\documentstyle[epsf]{europhys}

\def\etal{{\hbox{{\tenit\ et al.\/}\tenrm :\ }}}

\def\And{{\rm and\ }}

\newif\ifboo \boofalse

\begin{document}
\euro{}{}{}{}
\Date{}
\shorttitle{G.-S. Tian \etal Superconducting Correlations...}

\title
{Superconducting correlations in ultra-small metallic grains}
\author{Guang-Shan Tian\inst{1,2},
Lei-Han Tang\inst{1} \And Qing-Hu Chen\inst{1,3}}
\institute{
\inst{1} Department of Physics, Hong Kong Baptist University, 
Kowloon Tong, Hong Kong\\
\inst{2} Department of Physics, Peking University, Beijing 100871, China\\
\inst{3} Department of Physics, Zhejiang University, Hangzhou 310027, China
}
\rec{}{}

\pacs{
\Pacs{74}{20Fg}{BCS theory and its development}
\Pacs{71}{10$-$w}{Theories and models of many electron systems}
\Pacs{71}{24$+$q}{Electronic structure of clusters and nanoparticles}
}

\maketitle

\begin{abstract}
To describe the crossover from the bulk BCS
superconductivity to a fluctuation-dominated regime in ultrasmall
metallic grains, new order parameters and correlation functions,
such as ``parity gap'' and ``pair-mixing correlation function'',
have been recently introduced. In this paper, we discuss the
small-grain behaviour of the Penrose-Onsager-Yang off-diagonal
long-range order (ODLRO) parameter in a pseudo-spin representation.
Relations between the ODLRO parameter and those mentioned above
are established through analytical and numerical calculations.

\end{abstract}
The hallmarks of superconductivity --- zero resistance and the Meissner
effect --- both disappear when the size of a superconducting grain
shrinks to a few nanometers\cite{Anderson}. 
Naturally, one would like to know if there are
remnant effects of bulk superconductivity 
that differentiate normal and superconducting particles in this size range,
and if so, how these effects can be described at a quantitative level.
In a series of experiments on nanometer-sized Al grains,
Ralph, Black and Tinkham\cite{Ralph}
have shown that a gap in the excitation spectrum significantly
larger than the typical level spacing between single-electron
eigenstates still exists and is destroyed only by a sufficiently strong
magnetic field. The size of the energy gap depends on the parity of 
the electron number $N$ on the grain. 
These results indicate that pairing 
interactions remain important on nanometer scale.

Several theoretical approaches have been advanced to explain the
experimental findings mentioned above, and to address the broader
issue of superconductivity at nanometer scale
[3-14].
Attention has been focused on the applicability of
the Bardeen-Cooper-Schrieffer (BCS) theory\cite{Bardeen} to a situation where
(i) the average level spacing 
becomes comparable to the bulk superconducting gap and
(ii) the charging energy for adding an electron
becomes the largest energy scale and effectively fixes $N$.
The converging view is that, under condition (i), crossover takes
place from the BCS mean-field regime to a fluctuation-dominated regime,
though remnant pair correlations exist even in the latter case.
A general framework for addressing the fluctuation effects has been
proposed by Matveev and Larkin\cite{Matveev}. This is supplemented by
numerical calculations and exact solutions\cite{exact}
which offer a quantitative account of the crossover.

Although the problem of small-grain superconductivity under
the BCS Hamiltonian is essentially solved, 
one still owes for a simple viewpoint which is generalisable to 
situations where mathematical difficulties prevent an
explicit solution. The usual way for achieving such a goal is to 
identify a suitable order parameter which, in the present case, becomes
a nontrivial task due to condition (ii). Matveev and Larkin proposed to
use the ground state parity gap $\Delta_P=E_{N+1}-(E_{N}+E_{N+2})/2$ 
($N$ even) of an unpaired electron for this purpose.
An alternative suggestion was made in Refs.\cite{Mastellone,Braun,Braun1}
where a ``pair mixing parameter'' $\Delta_{\rm MIX}$ was introduced.
Although both quantities carry the effect of pair scattering,
they do not probe directly the pair coherence which is a defining 
property of the superconducting state or the remnant of it.

In this paper, we analyse the small-grain behaviour of yet another candidate, 
the Penrose-Onsager-Yang off-diagonal long-range order (ODLRO) 
parameter $\Delta_{\rm OD}$.\cite{Penrose,Yang} 
For the BCS Hamiltonian, $\Delta_{\rm OD}$ yields directly
the strength of a spontaneous symmetry-breaking field, whose
quantum fluctuations manifest in the ultra-small grain limit.
The crossover mentioned above is attributed to a
gradual weakening of this field against another energy scale of the 
problem, the level spacing $\delta$ at the Fermi surface.
Analytical and numerical results are presented to
relate $\Delta_{\rm OD}$ to $\Delta_{\rm MIX}$ and
$\Delta_P$.

Let us start with the BCS Hamiltonian for a metallic grain,
\begin{equation}
H_{\rm BCS} = \sum_{k,\sigma}
\epsilon_k c_{k\sigma}^\dagger c_{k\sigma} -
g \sum_{k, k'}c_{k\uparrow}^\dagger c_{k\downarrow}^\dagger
c_{k^\prime\downarrow} c_{k^\prime\uparrow}.
\label{BCS-Hamiltonian}
\end{equation}
Here $c_{k\sigma}^\dagger$ ($c_{k\sigma}$) are creation (annihilation)
operators of electrons in single-particle eigenstate $|k\rangle$
with spin $\sigma$ and energy $\epsilon_k$, and $g (>0)$ is the 
interaction constant. The sums in Eq. (\ref{BCS-Hamiltonian})
extend over states in the energy range
$-\hbar\omega_D\leq\epsilon_k\leq\hbar\omega_D$ around the Fermi level,
where $\omega_D$ is the Debye frequency.
To be definite, we take $\epsilon_k=k\delta$, where $\delta$
is the average level spacing, and $k$ is an integer
running from $-L$ to $L=\hbar\omega_D/\delta$.
The total number of levels involved, $\Omega=2L+1$, 
is proportional to the grain volume.
The only other dimensionless parameter of the model is the relative
interaction strength $\alpha=g/\delta$, which is expected to depend 
only weakly on grain size\cite{Braun1}. 

Under Eq. (\ref{BCS-Hamiltonian}), levels occupied by a single
electron are inert and do not participate in the pair
scattering process. For the remaining levels, the BCS Hamiltonian
can be written in a pseudo-spin form\cite{Anderson1},
\begin{equation}
H_{\rm BCS}=\sum_k\epsilon_k(1+2s_k^z)-g\sum_{k, k'}s_k^+s_{k'}^-,
\label{H-pseudo}
\end{equation}
where $s_k^+=c_{k\uparrow}^\dagger c_{k\downarrow}^\dagger$,
$s_k^-=c_{k\downarrow} c_{k\uparrow}$, and
$s_k^z=(c_{k\uparrow}^\dagger c_{k\uparrow}+
c_{k\downarrow}^\dagger c_{k\downarrow}-1)/2$
are spin-${1\over 2}$ operators.
In terms of the total spin operator ${\bf S}=\sum_k{\bf s}_k$,
the interaction term takes the form,
\begin{equation}
H_I=-gS^+S^-=-g{\bf S}^2+gS_z^2-gS_z.
\label{H-int}
\end{equation}
Thus the interaction energy $E_I$ is maximised by
the state with the largest total spin $S$ but the
smallest $S_z$ (in magnitude), i.e., the state with the largest XY component.
On the other hand, the single-particle energies
$\epsilon_k=k\delta, k=-L,\ldots, L,$ act as a nonuniform polarising
field in the $z$-direction.
Competition of these two tendencies determines the
ground state of the system.

In the usual mean-field analysis, $h_\pm\equiv gS^\pm$ 
is treated as a classical spontaneous symmetry-breaking
field coupled to the XY-component of the pseudo-spins.
The strength of this field is precisely the BCS gap parameter, i.e.,
$|h_\pm|\equiv g\langle S^+S^-\rangle^{1/2}=\Delta_{\rm BCS}$. 
The phase of $S^\pm$ is undefined for a system with a
fixed electron number $N$ which is an eigenstate of $S_z$\cite{Anderson2}.
As the size of the grain shrinks, quantum fluctuations of the
amplitude of $S^\pm$ become important when $h_\pm$ no longer
dominates over $\epsilon_k$ in the neighborhood of the Fermi level.

The Penrose-Onsager-Yang off-diagonal order parameter,
\begin{equation}
\Delta_{\rm OD} \equiv {1\over\Omega}
\langle\sum_{k, k'}
c_{k\uparrow}^\dagger c_{k\downarrow}^\dagger
c_{k'\downarrow} c_{k'\uparrow}\rangle,
\label{odop}
\end{equation}
offers a convenient measure of the strength of 
the spontaneous symmetry-breaking field 
in both canonical and grand-canonical
ensembles. Here $\langle\cdot\rangle$ denotes either the expectation value
over a pure state or an ensemble average.
It has been shown that, in the thermodynamic limit $\Omega\rightarrow\infty$, 
the existence of the ODLRO (in $k$-space) is equivalent to 
$\Delta_{\rm OD}$ being extensive in $\Omega$
\cite{Penrose,Yang,Tian2}. The pseudo-spin form of $\Delta_{\rm OD}$
is given by,
\begin{equation}
\Delta_{\rm OD}={1\over\Omega}\langle S^+S^-\rangle
\equiv {1\over\Omega}S_{\rm XY}^2,
\label{spin-od}
\end{equation}
where $S_{\rm XY}$ is an effective XY-spin.
In the bulk limit, $S_{\rm XY}$ is also extensive in $\Omega$
or the grain volume. The off-diagonal correlations of the Cooper
pairs are reflected in deviations from the free-electron results
$\Delta_{\rm OD}=1/2$ and $S_{\rm XY}=\sqrt{\Omega/2}$. 

We now consider the relation between $\Delta_{\rm OD}$ and the
pair-mixing parameter\cite{Delft,Mastellone,Braun,Braun1},
\begin{equation}
\Delta_{\rm MIX}\equiv \sum_k u_k v_k,
\label{pairing}
\end{equation}
where 
\begin{equation}
u_k=\langle c_{k\downarrow} c_{k\uparrow}
c_{k\uparrow}^\dagger c_{k\downarrow}^\dagger\rangle^{1/2}, 
\quad v_k=\langle c_{k\uparrow}^\dagger c_{k\downarrow}^\dagger
c_{k\downarrow} c_{k\uparrow}\rangle^{1/2},
\label{u_kv_k}
\end{equation}
are occupation amplitudes of paired holes and electrons, respectively.
In terms of the pseudo-spins, we have,
\begin{equation}
\Delta_{\rm MIX}=\sum_k \sqrt{{1\over 4}-\langle s_k^z\rangle^2}.
\label{spin-mix}
\end{equation}
Since pair-scattering yields partially occupied levels, 
$\Delta_{\rm MIX}$ assumes a nonvanishing value under 
(\ref{BCS-Hamiltonian}). Unlike $\Delta_{\rm OD}$, however,
$\Delta_{\rm MIX}$ does not differentiate between coherent 
and incoherent scattering. The following theorem 
relates $\Delta_{\rm MIX}$ to $\Delta_{\rm OD}$.

{\bf Theorem:} For the global ground state $\Psi_0$ of the BCS
Hamiltonian (\ref{BCS-Hamiltonian}) with $P$ Cooper pairs,
the following inequalities hold,
\begin{equation}
\frac{\Delta_{\rm MIX}^2}{\Omega}
\leq \Delta_{\rm OD}
\leq {P\over\Omega}+\Delta_{\rm MIX}\Bigl(1-{1\over \Omega}\Bigr).
\label{Inequality1}
\end{equation}

{\it Proof:} We start with the inequality,
\begin{equation}
\Delta_{\rm OD}
\leq {P\over\Omega}+
\frac{1}{\Omega} \sum_{k\ne k^\prime} 
\big|\langle c_{k\uparrow}^\dagger c_{k\downarrow}^\dagger
c_{k^\prime\downarrow} c_{k^\prime\uparrow}\rangle\big|. 
\label{Definition1}
\end{equation}
An upper bound to each term in the sum is obtained
by inserting the identity operator
$\hat{I} \equiv \sum |\Psi_h\rangle \langle\Psi_h|$
between the pair creation and annihilation operators and
applying the Cauchy-Schwartz inequality
$|\sum_n a_nb_n|^2\leq(\sum_n|a_n|^2)(\sum_n |b_n|^2)$, 
\begin{eqnarray}
\big|\langle c_{k\uparrow}^\dagger c_{k\downarrow}^\dagger
c_{k^\prime\downarrow} c_{k^\prime\uparrow}\rangle\big|
&\le&
\sqrt{\langle c_{k\uparrow}^\dagger
c_{k\downarrow}^\dagger c_{k\downarrow}c_{k\uparrow} \rangle} 
\sqrt{\langle c_{k^\prime\uparrow}^\dagger
c_{k^\prime\downarrow}^\dagger c_{k^\prime\downarrow}
c_{k^\prime\uparrow}\rangle} \nonumber\\
&=&v_k v_{k^\prime}.
\label{Bound1}
\end{eqnarray}
Similarly, we can show that
\begin{equation}
\big|\langle c_{k^\prime\downarrow} c_{k^\prime\uparrow}
c_{k\uparrow}^\dagger c_{k\downarrow}^\dagger\rangle \big|\le
u_{k^\prime} u_k.
\label{Bound2}
\end{equation}
Multiplying (\ref{Bound1}) with (\ref{Bound2}), we obtain,
\begin{equation}
\big|\langle
c_{k\uparrow}^\dagger c_{k\downarrow}^\dagger
c_{k^\prime\downarrow} c_{k^\prime\uparrow}\rangle \big|\le 
\sqrt{u_k v_k} \sqrt{u_{k^\prime} v_{k^\prime}}.
\label{Bound3}
\end{equation}
Substitution of (\ref{Bound3}) into (\ref{Definition1}) yields
\begin{equation}
\Delta_{\rm OD} \le {P\over\Omega}+
\frac{1}{\Omega} \Bigl(\sum_k \sqrt{u_k v_k}\Bigr)^2 -
\frac{\Delta_{\rm MIX}}{\Omega}.
\label{Bound4}
\end{equation}
The upper bound to $\Delta_{\rm OD}$ in (\ref{Inequality1})
is obtained by applying again the Cauchy-Schwartz inequality to
the sum in (\ref{Bound4}) with $a_k=1$ and $b_k=\sqrt{u_kv_k}$.

Unlike the upper bound, the lower bound in (\ref{Inequality1})
makes specific reference to the BCS Hamiltonian for which
the interaction energy $E_I = -g \Omega \Delta_{\rm OD}$,
and to the global ground-state $\Psi_0$.
Leaving out levels occupied by a single electron,
we may write (\ref{u_kv_k}) as,
\begin{equation}
u_k=\langle\Psi_0|1 - n_k|\Psi_0\rangle^{1/2},\quad
v_k=\langle\Psi_0|n_k|\Psi_0\rangle^{1/2}, 
\label{uv1}
\end{equation}
where $n_k$ is the number operator for a pair.
Consider now a normalised BCS-like wave function,
\begin{equation}
\widetilde{\Psi}_0 \equiv
\prod_k \left(u_k + v_k c_{k\uparrow}^\dagger
c_{k\downarrow}^\dagger \right)|0\rangle.
\label{BCS_Function}
\end{equation}
The expectation values of $H_{\rm BCS}$ in the two states
$\Psi_0$ and $\widetilde\Psi_0$ are given by,
\begin{eqnarray}
\langle\Psi_0|H_{\rm BCS}|\Psi_0\rangle & = & E_K + E_I,
\label{gs_energy}\\
\langle\widetilde{\Psi}_0|H_{\rm BCS}|\widetilde{\Psi}_0\rangle
& = & E_K - g \Delta_{\rm MIX}^2
- g \sum_k v_k^4,
\label{trial_energy}
\end{eqnarray}
where $E_K=\sum_k 2\epsilon_k v_k^2$ is the kinetic energy
of pairs in the ground state.
The lower bound in (\ref{Inequality1}) is obtained by noting that
(\ref{trial_energy}) is an upper bound to (\ref{gs_energy}).
{\bf QED}.

We note in passing that the particle-hole symmetry of $H_{\rm BCS}$
suggests that the global ground state (for a fixed parity)
is achieved at the half-filling value $P=\Omega/2$,
though a rigorous proof is so far lacking.
By the very construction of the BCS Hamiltonian (\ref{BCS-Hamiltonian}),
the global ground state (with respect to all possible values of $P$)
is the one physically realised.

{\it A direct corollary of the theorem is that, for the
ground state of the BCS Hamiltonian, $\Delta_{\rm MIX}$
is extensive if and only if $\Delta_{\rm OD}$ is extensive}.

In Ref.\cite{Mastellone}, it was suggested that $\Delta_{\rm MIX}$
may become non-extensive when $\alpha=g/\delta$ is less than
a certain critical value $\alpha_c>0$. This is in apparent contradiction
with the well-known result that the ground state 
of the BCS Hamiltonian has ODLRO for any $\alpha>0$.
Since the BCS ground state energy $E_{\rm BCS}$ provides
another upper bound to the true global ground state energy $E_0$,
it is possible to show that
\begin{equation}
\Delta_{\rm OD}\geq -E_{C,\rm BCS}/(g\Omega),
\label{ineq3}
\end{equation}
where $E_{C,\rm BCS}$ is the BCS condensation energy which,
for large $\Omega$, is given by,
\begin{equation}
E_{C,\rm BCS}=-{\Delta_{\rm BCS}^2\over\delta}
\Bigl[1+\sqrt{1+\bigl({\Delta_{\rm BCS}\over L\delta}\bigr)^2}\,\Bigr]^{-1}.
\label{bulk-cond}
\end{equation}
Here $\Delta_{\rm BCS} =L\delta/\sinh(\delta/g)$ is the BCS gap parameter.
Since $E_{C,\rm BCS}$ is extensive, both $\Delta_{\rm OD}$ and
$\Delta_{\rm MIX}$ are extensive in the bulk limit.

Let us now turn to the relations among the spontaneous
symmetry breaking field $h_\pm$, the parity parameter $\Delta_P$,
and other spectroscopic parameters.
In the bulk limit, $|h_\pm|\gg\delta$, so that pseudo-spins around
the Fermi level are completely aligned with the symmetry-breaking field.
The parity gap $\Delta_P$ is simply the energy cost for removing
a level at the Fermi energy. This has the effect of reducing
$S_{\rm XY}$ by $1/2$ which, from Eq. (\ref{H-int}),
carries an energy cost $gS_{\rm XY}=|h_\pm|=\Delta_{\rm BCS}$.
Likewise, the energy change for reversing the pseudo-spin, which 
reduces $S_{\rm XY}$
by one, is $2gS_{\rm XY}=2\Delta_{\rm BCS}$. This is the energy gap
for exciting a pair. The energy gap for breaking a pair can be
derived in a similar fashion and is again given by $2\Delta_{\rm BCS}$
in this limit. It is interesting to explore how these results
change for ultra-small grains.

One possible scheme for addressing pair-scattering effects
in the fluctuating regime is a renormalised perturbation theory
advanced by Matveev and Larkin\cite{Matveev}.
Treating the interaction term as a perturbation and computing
the ground state energies to the second order in $g$, Matveev and
Larkin obtained,
\begin{equation}
\Delta_P\simeq {g\over 2}+{g^2\over 2\delta}\ln{\Omega\over 2}.
\label{larkin}
\end{equation}
The second term in (\ref{larkin}) can be absorbed in the form of
a renormalised coupling constant $\tilde g=g/[1-(g/\delta)\ln(\Omega/2)]$,
and finally, $\Delta_P$ can be expressed in a scaling form,
\begin{equation}
\Delta_P=\Delta_{\rm BCS}f(\delta/\Delta_{\rm BCS}),
\label{scaling}
\end{equation}
where $f(x)\simeq 1$ for $x\ll 1$ and $f(x)\simeq x/(2\ln x)$ for $x\gg 1$.
Although the scaling ansatz (\ref{scaling}) is supported by
numerical results, a full-fledged renormalisation group derivation
is lacking.

To check if a similar scaling form applies to the spontaneous
symmetry-breaking field $h_\pm$, we have carried out
exact diagonalisation of $H_{\rm BCS}$ 
for small systems ($\Omega=7,9,\ldots,21$)
using the Lanczos method. 
Figure 1(a) shows $gS_{\rm XY}/\Delta_{\rm BCS}$ (diamonds)
versus $\delta/\Delta_{\rm BCS}$ for a range of $\alpha$ and 
$\Omega$ values.
For comparison, we have also included data for the
parity parameter $\Delta_P$ (open circles) and the pair-mixing parameter
$\Delta_{\rm MIX}$ (solid circles) in a suitably scaled form.
The scaling ansatz (\ref{scaling}), which describes well 
the crossover behaviour of the parity gap $\Delta_P$,
does not seem to apply to $gS_{\rm XY}$ and $g\Delta_{\rm MIX}$
in the fluctuation-dominated regime. The lower-bound in
(\ref{Inequality1}) (equivalent to $\Delta_{\rm MIX}\leq S_{\rm XY}$)
is obviously satisfied by the data.
Note that, in the small grain limit, $S_{\rm XY}$ is dominated by
zero-point fluctuations of $S^\pm$ rather than the spontaneous
symmetry-breaking effect. This is reflected in the increase of
$gS_{\rm XY}$ from its bulk value $\Delta_{\rm BCS}$ seen 
in Fig. 1(a).

By computing the ground state wave function to the first order in $g$,
we derived the following result,
\begin{equation}
\Delta_{\rm OD}= {1\over 2}+{g\over\delta}\ln 2+O(g^2).
\label{Delta-pert}
\end{equation}
Comparison with Eq. (\ref{larkin}) suggests, on the strongly-fluctuating
side,
\begin{equation}
\Delta_{\rm OD}\simeq {1\over 2}+(2\ln 2){\Delta_P\over\delta}.
\label{delta-rg}
\end{equation}
Figure 1(b) shows $\Delta_{\rm OD}$ against $\Delta_P/\delta$
for different values of $\alpha=g/\delta$. At the lower-left
corner of the graph, there appears to be a good data collapse
which compares favorably with Eq. (\ref{delta-rg})
as indicated by the solid line. (Note that part of the descrepancy
is due to $P/\Omega$ being slightly less than $1/2$ for odd $\Omega$.)
On the other hand, in the bulk limit, 
$\Delta_{\rm OD}=\Omega^{-1}(\Delta_{\rm BCS}/g)^2$
can not be expressed as a function of $\Delta_P/\delta$ only.
This again indicates that the fluctuation and correlation effects can not be
fully captured by a one-parameter renormalisation group theory.

In summary, we have established some inequalities that relate
the Penrose-Onsager-Yang off-diagonal order parameter $\Delta_{\rm OD}$ to
the pair-mixing parameter $\Delta_{\rm MIX}$ recently proposed
to characterise the crossover from the bulk BCS regime to
the fluctuation-dominated regime. It is shown that, for the ground state
of the BCS Hamiltonian, both quantities are extensive in the bulk limit.
However, $\Delta_{\rm OD}$ offers a better measure of the superconducting
condensation in general as it is sensitive also to the ``phase''
of electron pairs. The nature of the spontaneous symmetry breaking for
superconducting grains at a fixed electron number $N$,
and relationships among various excitations,
are examined in the framework of the pseudo-spin representation.
Our exact numerical results for small systems support the
previous view that the fluctuation-dominated quantum regime is reached when
the BCS gap parameter $\Delta_{\rm BCS}$ becomes
comparable to the level spacing $\delta$.

{\bf Acknowledgments:} This work is supported by
the Hong Kong Baptist University under Grant FRG/97-98/II-78,
and by the Chinese National Science Foundation under grant No. 19874004.


\newpage
\begin{figure}
\epsfxsize=\linewidth
\epsffile{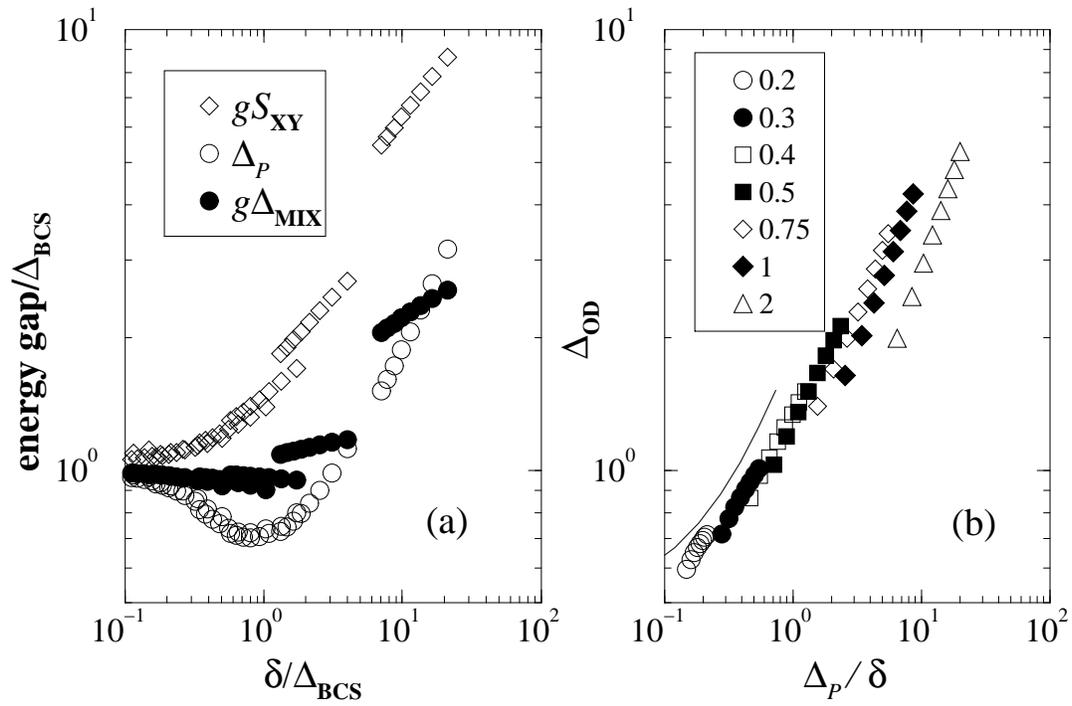}
\caption{(a) The spontaneous symmetry-breaking
field $gS_{\rm XY}$ ($\diamond$), parity parameter ($\circ$), 
and pair-mixing parameter $\Delta_{\rm MIX}$
($\bullet$) against $\delta/\Delta_{\rm BCS}$ in suitably scaled form. 
(b) The ODLRO parameter plotted against $\Delta_P/\delta$. Solid line 
indicates result from the perturbative calculation.
Numbers in legend give the values of $\alpha=g/\delta$ for each
data set.
}
\label{fig1}
\end{figure}

\end{document}